# PlugSelect: Pruning Channels with Plug-and-Play Flexibility for Electroencephalography-based Brain Computer Interface

Xue Yuan, Keren Shi, Ning Jiang, *Senior Member*, *IEEE*, Jiayuan He, *Member*, *IEEE*

***Abstract*—Automatic minimization and optimization of the number of the electrodes is essential for the practical application of electroencephalography (EEG)-based brain computer interface (BCI). Previous methods typically require additional training costs or rely on prior knowledge assumptions. This study proposed a novel channel pruning model, plug-and-select (PlugSelect), applicable across a broad range of BCI paradigms with no additional training cost and plug-and-play functionality. It integrates gradients along the input path to globally infer the causal relationships between input channels and outputs, and ranks the contribution sequences to identify the most highly attributed channels. The results showed that for three BCI paradigms, *i.e.*, auditory attention decoding (AAD), motor imagery (MI), affective computation (AC), PlugSelect could reduce the number of channels by at least half while effectively maintaining decoding performance and improving efficiency. The outcome benefits the design of wearable EEG-based devices, facilitating the practical application of BCI technology.**

## I. INTRODUCTION

Electroencephalography (EEG) is the electrical manifestation of brain activities and contain neural information of the central nervous system (CNS) [1]. It is commonly employed as source signal for the application of brain computer interface (BCI), which provides the possibility of directly controlling machines based on the human intentions decoded from neural signals [2]. There are multiple types of EEG-based BCI paradigms, such as auditory attention decoding (AAD), motor imagery (MI), affective computation (AC), etc. AAD is a task which is designed to decode which voice the participant is listening from EEG signals in a multi-speaker condition [3]. MI is to decode the type of the imagined movements from EEG signals [4]. AC is to obtain the emotion of the participant from EEG signals [5]. It is suggested that multiple types of information could be decoded from EEG signals. The EEG-based BCI is a practical noninvasive technology of restoring and augmenting the functions of the humans.

Though promising, there were several challenges for practical use of EEG-based BCI system. One key challenge involves the automatic optimization of the number of the electrodes, or channels, placed on the scalp. For most current EEG tasks, tens of the electrodes are employed for achieving high decoding performance. However, excessive electrodes increased the complexity and decreased the portability of the system. As such, channel pruning, *i.e.*, automatic minimization and optimization of the number of the electrodes, is crucial for the practical application of the system [6], [7], [8].

The optimal electrode placement is subject to the BCI paradigm, and the number of the electrodes required. Current EEG channel pruning approaches mainly include end-to-end deep learning (DL) algorithms, machine learning methods, and statistical techniques [8]. Iteratively applying DL classifiers to eliminate channels with minimal contribution to the decoding performance for a specific task is the most direct approach for channel selection. Kashefi et al. applied shallow convolutional neural networks (CNNs) iteratively to perform channel selection for MI [8], while Mirkovic et al. set up an offline iterative process for channel selection in auditory attention tasks [9]. In addition, Xu et al. proposed an end-to-end DL model based on a weighted residual structure to identify a subset of invariant channels relevant to auditory tasks at the group level [10]. Sun et al. incorporated a channel selection module into an end-to-end MI decoding model [11]. However, channel selection schemes that rely on iterative classifiers and training processes not only increase training and learning costs but also complicate physiological interpretation. Similar to iterative processes, Narayanan et al. employed a greedy algorithm to search for channels relevant to auditory attention [12]. Moreover, some researchers have utilized the weights of intermediate layers in DL algorithms for channel selection, thereby reducing computational costs. Wang et al. used the 2-norm of the spatial convolution layer weights to determine the MI classification contribution [7], Lin et al. considered attention scores as indicators of relevance for AC tasks [13], and Cai et al. attempted to quantify the importance of EEG channels for AAD tasks using graph attention weights [14]. However, the weights of a given layer may not directly reflect the causal relationship between input and output and often

This work was supported by the Fundamental Research Funds for the Central Universities under Grant YJ202373, Science and Technology Major Project of Tibetan Autonomous Region of China under Grant XZ202201ZD0001G, 1.3.5 project for Disciplines of 1435 Excellence Grant from West China Hospital under Grant ZYYC22001, and key project from Med-X Center for Manufacturing under Grant 0040206107007.
(Corresponding author: Jiayuan He; Email: jiayuan.he@wchscu.cn).

Xue Yuan, Keren Shi, Ning Jiang, and Jiayuan He are with National Clinical Research Center for Geriatrics, West China Hospital, Sichuan University, Chengdu, Sichuan 610017, China, and also with Med-X Center for Manufacturing, Sichuan University, Chengdu, Sichuan 610017, China.

exhibit poor stability. Therefore, some researchers have adopted machine learning or statistical methods that more directly capture causal relationships for channel selection. Wang et al. [15] selected four optimal channels using weight vectors derived from common spatial patterns (CSP), while Yong et al. further enhanced the sparsity of CSP weight vectors by introducing ℓ1-norm regularization [16], [17]. Li et al. selected channels by calculating the statistical significance of power spectral parameters in relation to the task [18], and Jing et al. proposed a correlation-based channel selection method [19], [20]. These methods typically rely on linear assumptions, which may fail to fully capture the complex nonlinear features present in EEG signals. Furthermore, all of the aforementioned methods are tailored to specific BCI paradigms, and their portability and effectiveness across different platforms remain uncertain.

To address these challenges, we propose plug-and-select (PlugSelect), a framework for channel selection based on outcome attribution designed for the EEG-based BCI paradigms. PlugSelect could work with different neural networks and requires no additional training costs, prior knowledge, or assumptions for channel selection. Specifically, it consists of two modules: integrated gradients (IG) and ranking strategy (RS). (1) IG integrates the gradients along the input path to assess the global contribution of input channels to the prediction outcome, thereby providing a direct interpretation of the decision-making process in complex neural networks. This approach addresses the issue of traditional weight-based schemes' inability to stably capture the causal relationship between input and output, all while requiring no additional training costs. (2) RS, building upon the personalized channel selection scheme provided by IG, introduces a task-level channel ranking strategy aimed at addressing subject heterogeneity, thereby achieving a more stable and reliable subset of channels.

In addition, for verifying the plug-and-play functionality and broad effectiveness of the framework, three different types of EEG-based BCI paradigms were applied: 1) the new cue-masked AAD task, of which one is the two-class orientation attention (OA) decoding, and another is the two-class timbre attention (TA) decoding; 2) the traditional four-class MI task; 3) the three-class AC task. Using the PlugSelect framework, we then calculated the average classification contribution of each channel from the input multichannel EEG signal assessing its impact on prediction outcomes. And evaluated the model's performance under varying channel densities using metrics such as accuracy (ACC). Ultimately, we aimed to balance the number of channels, model decoding performance, and computational efficiency (number of samples processed per second) by selecting a task-relevant subset of channels. This provides guidance for designing more efficient, low-channel EEG caps suitable for different EEG-based BCI paradigms.

In conclusion, our main contributions are: (1) We propose PlugSelect as a plug-and-play systematic channel selection framework, which has been efficiently ported and validated across multiple data platforms (MI task and AC task). (2) PlugSelect preserves decoding efficiency while significantly reducing redundant channels. (3) 15 AAD-related channels are identified, and the channel pruning results demonstrate a strong correlation with downstream task performance.

## II Method

### A. Channel Pruning Framework

The proposed PlugSelect consists of two key modules: IG and RS, as shown in Fig. 1. The IG is designed to estimate the contribution of each channel for the downstream BCI tasks. The RS is used to process the difference among the subjects to mitigate the effects of subject variability and make the results generalized. The details of the two modules are described as follows:

*a). IG*

For a specific BCI task, IG employs linear interpolation and gradient summation along the path between a reference baseline and the EEG data to measure the contribution of each channel. Its input is raw EEG data, denoted as $\varphi \in \mathbb{R}^{C\times T}$ with the number of the channels as $C$ and the sample points in the $s^{th}$ decision window as $T$, and a pre-trained neural network model for the downstream BCI tasks with full channels, denoted as $\Delta_x$. For the $i^{th}$ channel, the attribution $\phi_i^s$ at the $s^{th}$ decision window is calculated as follows:

$$\phi_i^s(h, \varphi_i, \varphi_i') = (\varphi_i - \varphi_i') \times \mathcal{G} \times \frac{1}{M} \quad (1)$$

where $\varphi_i'$ is the reference baseline for the data of the $i^{th}$ channel $\varphi_i \in \mathbb{R}^{1\times T}$ with a zero-baseline in this study. $M$ represents the number of steps in the Riemannian path integral. $\mathcal{G}$ denotes the gradient integral over the path, which is defined as follows:

$$\mathcal{G} = \sum_{m=1}^{M} \frac{\partial\left(h(\rho_m, \Delta_x)\right)}{\partial \varphi_i} \quad (2)$$

$$\rho_m = \varphi_i^{'} + \frac{m}{M} \times \left(\varphi_i - \varphi_i^{'}\right) \quad (3)$$

where $h(\cdot)$ denotes the mapping function of the pre-trained model $\Delta_x$, which is responsible for transforming the input $\rho_m$ into a specific representation in the output space. $\rho_m$ is a scaling variable at step $m$. IG employs an integral method to express the impact of each incremental change from the zero-baseline to the complete input on the output in a fine-grained manner, rather than merely focusing on the singular effect of the complete input on the model output. This makes it apply to the nonlinear model. Additionally, IG integrates the cumulative contribution along the entire path from the reference baseline to the actual input, providing guidance in the channel pruning process by retaining key channels and pruning redundant ones.

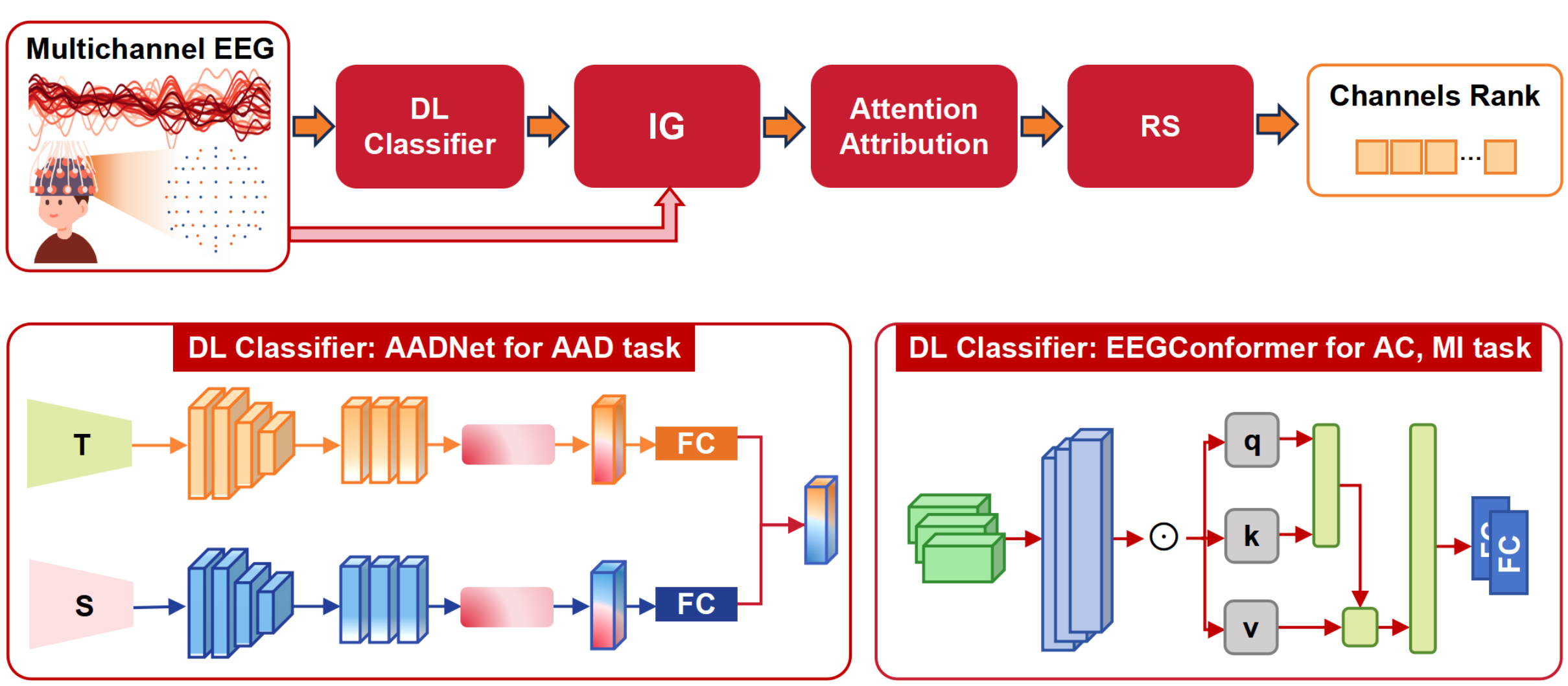

Fig. 1. The framework of PlugSelect, contains two modules: IG and RS. IG, integrated gradients. RS, ranking strategy. IG evaluates the global contribution of input channels to the prediction outcome, providing a direct interpretation of the decision-making process in neural networks. RS, building upon the personalized channel selection provided by IG, introduces a task-level channel ranking strategy. Together, these modules enable PlugSelect to perform efficient and interpretable channel selection for various EEG-based BCI tasks without requiring additional training costs or prior assumptions.

Suppose there are $n$ decision windows for each subject, the contribution of the $i^{th}$ channel is the summation of the contributions across all the windows, which is calculated as follows:

$$\phi_i = \sum_{s=1}^{n} \phi_i^s \quad (4)$$

The contribution value ranged from -1 to 1. If it is greater than zero, it indicates that the channel positively contributes to the identification of the downstream task. If it is less than zero, it indicates a negative contribution. The larger the absolute value of IG, the stronger the channel's influence on the model's classification.

Algorithm 1 shows the pseudo-code for the reasoning component of the proposed PlugSelect method.

*b). RS*

As the results of channel selection might be varied among the subjects, two strategies, *i.e.*, averaging and voting, were employed on the results of the subjects available for generalization. For averaging, it considers shared statistics. For each channel, attribution of each subject from IG were averaged, and then the average values were ranked from highest to lowest for channel selection. For voting, it considers sample specificity. For each subject, channel attribution from IG is ranked, and then the occurrence of each channel in the top rank were calculated with the data of all the subjects. The electrodes with high occurrence were selected.

**Algorithm 1** IG.

**Input:** Full-channel EEG signals $\varphi \in \mathbb{R}^{C\times T}$, zero-baseline
**begin**
  Load the pretrained model $\Delta_x$;
  Compute $d$, the difference between $\varphi$ and baseline $\varphi'$
  **assign** $\mathcal{G} = 0$
  **for** each path step $m$ from 1 to $M$ **do**
    Compute $\rho_m$, the scaled input
      $\rho_m = \varphi' + \frac{m}{M} \times d$
    Compute $\mathcal{G}$, gradient of the model's output with $\rho_m$
      $\mathcal{G}' = \frac{\partial(h(\rho_m, \Delta_x))}{\partial \varphi}$
    Accumulate the gradients, $\mathcal{G} \mathrel{+}= \mathcal{G}'$
  $\phi_d = d \times (\frac{\mathcal{G}}{M})$, $\phi_d \in \mathbb{R}^{C\times T}$
  Compute all channel's attribution $\phi$
  $\phi = mean(\phi_d, axis = 1)$
  **return** $\phi$
**end**

*B. BCI Paradigms*

For investigating the feasibility of the proposed framework, three BCI paradigms were employed in this study, including AAD, MI, and AC. The dataset of AAD was self-collected, and the datasets of the other two paradigms were from the open data repository, which was BCI Competition IV-2a for MI and SEED for AC. The three EEG datasets varied in acquisition devices, experimental paradigms, subject numbers, and sample sizes, providing a comprehensive basis for fairly validating the multi-platform portability and effectiveness of the proposed method. Due to the difference in the number of channels among the datasets, the electrode positions from the 64-lead system were used as the foundational canvas for the analysis, as shown in Fig. 2.

*a) AAD*

The goal of AAD was to decode auditory attention from EEG. There were 30 healthy right-handed participants (aged 17-32 years, 15 females and 15 males with subject IDs S1 to S30) recruited in this study. Before the experiment, the procedures were provided and written informed consent was obtained from all the participants. The experimental protocol was in accordance with the Declaration of Helsinki, and approved by the Research Ethics Committee of West China Hospital, Sichuan University (# 2024582).

The experimental protocol was illustrated in Fig. 3. The experiment was conducted in a soundproof room where the participants' field of vision was limited to white walls. During the experiment, the participants were exposed to mixed male and female audio stimuli, and the EEG signals were recorded with a commercial device (Enobio EEG systems, NE Neuroelectrics, Spain). Based on the international 10-20 system, 32 electrode positions were selected across the entire scalp, *i.e.*, P8, T8, CP6, FC6, F8, F4, C4, P4, AF4, Fp2, Fp1, AF3, Fz, FC2, Cz, CP2, PO3, O1, Oz, O2, PO4, Pz, CP1, FC1, P3, C3, F3, F7, FC5, CP5, T7, P7. The sampling frequency was 500 Hz.

There were two AAD tasks, OA and TA. In the mixed speech stimulus trials, the target timbre and spatial location were randomized, but the number of trials was nearly equal. Each participant completed 12 mixed speech stimulus trials, with each trial having a stimulus duration of 70 seconds.

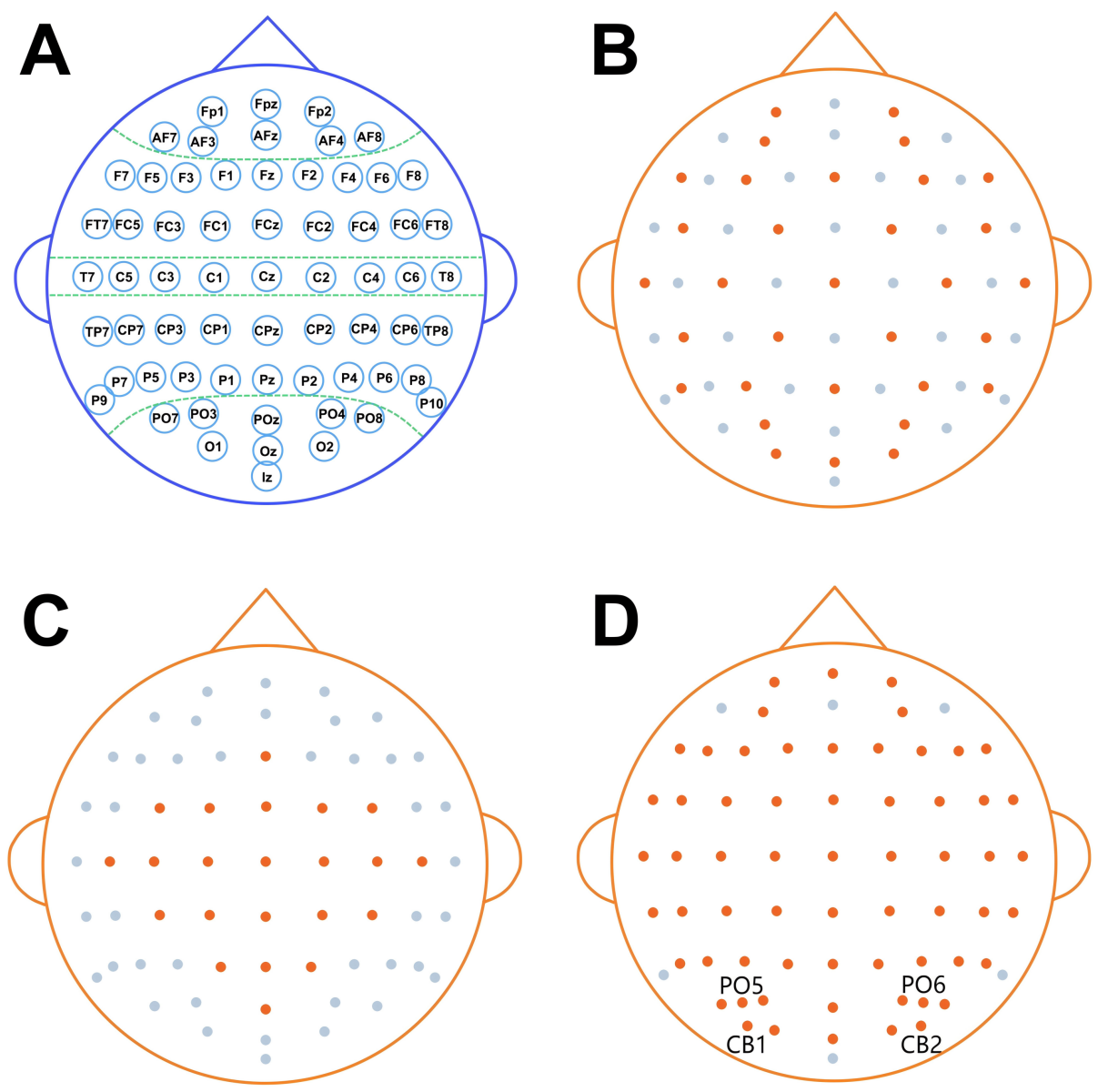

Fig. 2. Electrode position. (A) 64-lead position based on 10-20 system placement. (B) 32 electrode positions selected in AAD. (C) 22 electrode positions selected in MI. (D) 62 electrode positions selected in AC. Orange indicates selected electrode positions for each task.

For signal preprocessing, the collected data underwent successively average referencing and bandpass filtering (0.4-32 Hz). Then independent component analysis (ICA) was employed to eliminate artifacts such as eye movements and muscle activities. The data was segmented into 0.5-second windows with no overlap, resulting in 1,656 windows per participant.

*b) MI*

The goal of MI was to decode the imagined movements from EEG. An open dataset, BCI Competition IV-2a dataset provided by Graz University of Technology, was adopted in this study. It contains EEG data from 9 subjects. The paradigm involved MI tasks for four movement categories: left hand (L), right hand (R), feet (F), and tongue (T), as illustrated in Fig. 3 (C). Each subject completed two sessions on separate days, with each session comprising 72 EEG trials for each of the four MI tasks, recorded at a sampling rate of 250 Hz. A total of 22 electrodes, placed according to the 10-20 system, were used: Fz, FC3, FC1, FCz, FC2, FC4, C5, C3, C1, Cz, C2, C4, C6, CP3, CP1, CPz, CP2, CP4, P1, Poz, Pz, and P2. Our analysis focused on the 0-4 second window after cue onset, corresponding to [2, 6] seconds per trial.

For signal preprocessing, a band-pass filter was applied to the EEG data in the [4, 40] Hz range, as described in [21]. In this study, a 6th-order Chebyshev filter was employed to preserve task-relevant rhythms. However, the time-consuming nature of MI-EEG acquisition and the limited size of the dataset increase the risk of underfitting. As such, following previous studies [22], [23], the strategy of segmentation and reconstruction (S&R) in the time domain was adopted to generate additional data. Additionally, Z-score normalization was performed on the EEG data from Datasets MI, AC to mitigate fluctuations and non-stationarity. The normalization is as follows:

$$\bar{X} = \frac{X - \mu}{\sigma} \tag{5}$$

where $\mu$ and $\sigma$ denote the mean and standard deviation of the training set, respectively.

*c) AC*

The goal of AC was to decode emotion from EEG. An open dataset, SEED provided by Shanghai Jiao Tong University, was adopted in this study. It contains emotion-based EEG signals from 15 subjects. Each session involved 15 movie clips designed to evoke positive, neutral, and negative moods, with the paradigm illustrated in Fig. 3 (D). Data were collected across three sessions, spaced approximately one week apart. EEG signals were recorded from 62 electrodes at a sampling rate of 1000 Hz, and subsequently down sampled to 200 Hz. The 62 channels, placed according to the 10-20 system, included: Fp1, Fpz, Fp2, AF3, AF4, F7, F5, F3, F1, Fz, F2, F4, F6, F8, FT7, FC5, FC3, FC1, FCz, FC2, FC4, FC6, FT8, T7, C5, C3, C1, Cz, C2, C4, C6, T8, TP7, CP5, CP3, CP1, CPz, CP2, CP4, CP6, TP8, P7, P5, P3, P1, Pz, P2, P4, P6, P8, PO7, PO5, PO3, Poz, PO4, PO6, PO8, CB1, O1, Oz, O2, and CB2. Experiments were conducted based on the shortest trial length, consisting of 37,000 sample points (185 seconds).

For preprocessing, we applied a 6th-order Chebyshev band-pass filter to the data in the [4, 47] Hz range. Each sample was then segmented into non-overlapping one-second time windows, resulting in 2775 trials from a single session. Z-score normalization was adopted to mitigate fluctuations and non-stationarity, which was the same as MI.

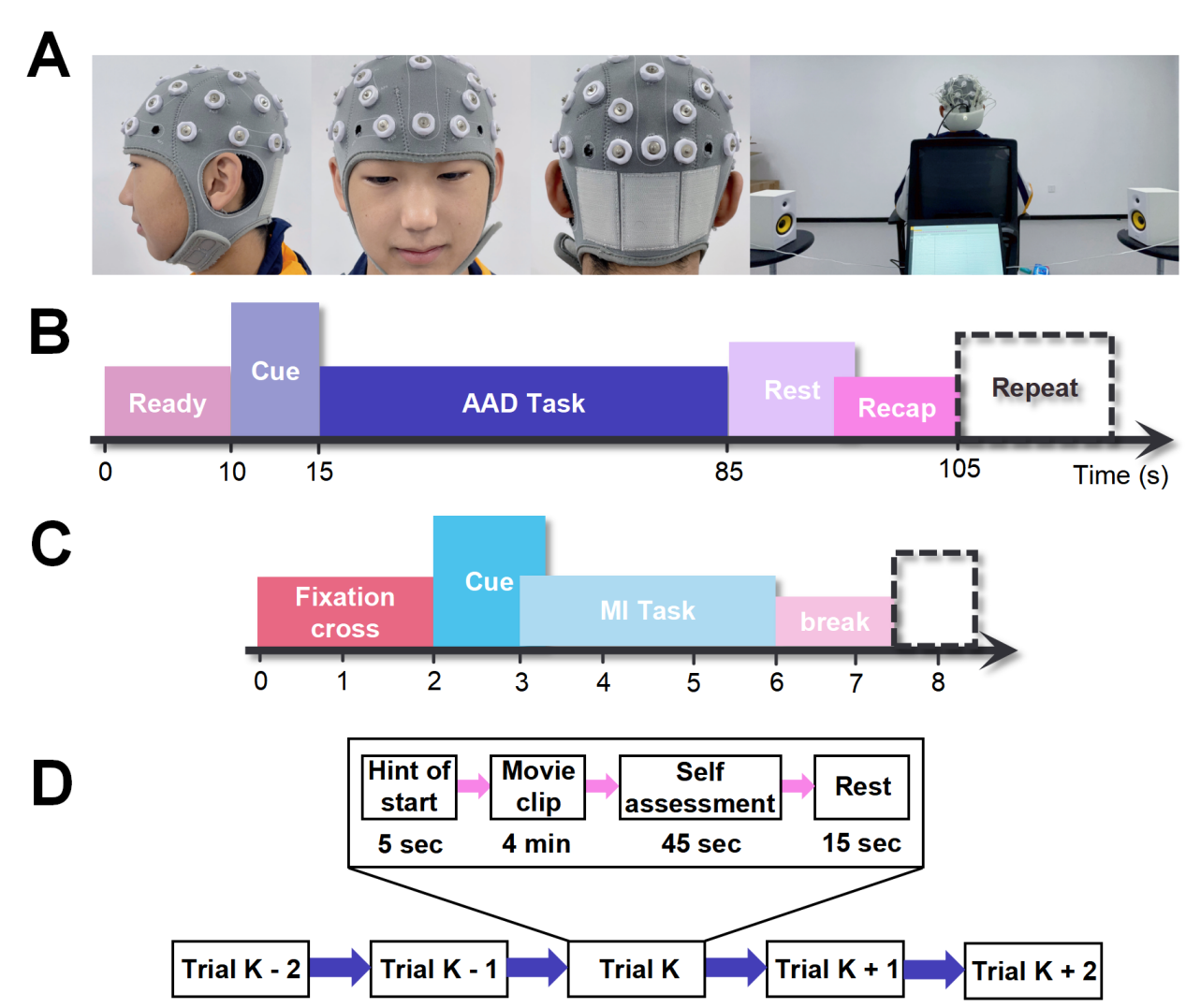

Fig. 3. Experimental paradigms for each task. (A) Experimental conditions for the AAD, illustrating the manner in which participants focus their attention and the presentation of auditory stimuli during the task. Paradigms for (B) AAD, (C) MI, (D) AC. Each experimental paradigm provides a detailed explanation of the task design and the presentation of stimuli aimed at enhancing the decoding of brain signals in different experimental contexts.

### C. Neural Network

The implementation of the proposed framework needed a pre-trained neural network with effective decoding capability.

In this study, the decoding model for the AAD task is AADNet[24], which demonstrated high classification accuracy for AAD tasks, as shown in Fig. 4. It comprises three main blocks: temporal convolution, spatial convolution, and hybrid decoding modules. The number of 2D convolution kernels for the three blocks is set to [32, 64, 64], with kernel sizes of [(1, 64), (32, 1), (1, 16)], and the number of kernels is kept consistent within each block.

For MI and AC, EEGConformer was adopted as the decoding model for the good performance in [23]. EEGConformer consists of two main modules: the spatio-temporal convolution module and the multi-head self-attention module. The spatio-temporal convolution module contains 40 2D convolution kernels, with kernel sizes of [(1, 25), ( $ch$ , 1)], where $ch$ represents the number of channels for the corresponding task. Both the MI task and the AC task execute the multi-head attention module six times, but the number of attention heads differs, which is 10 and 5, respectively.

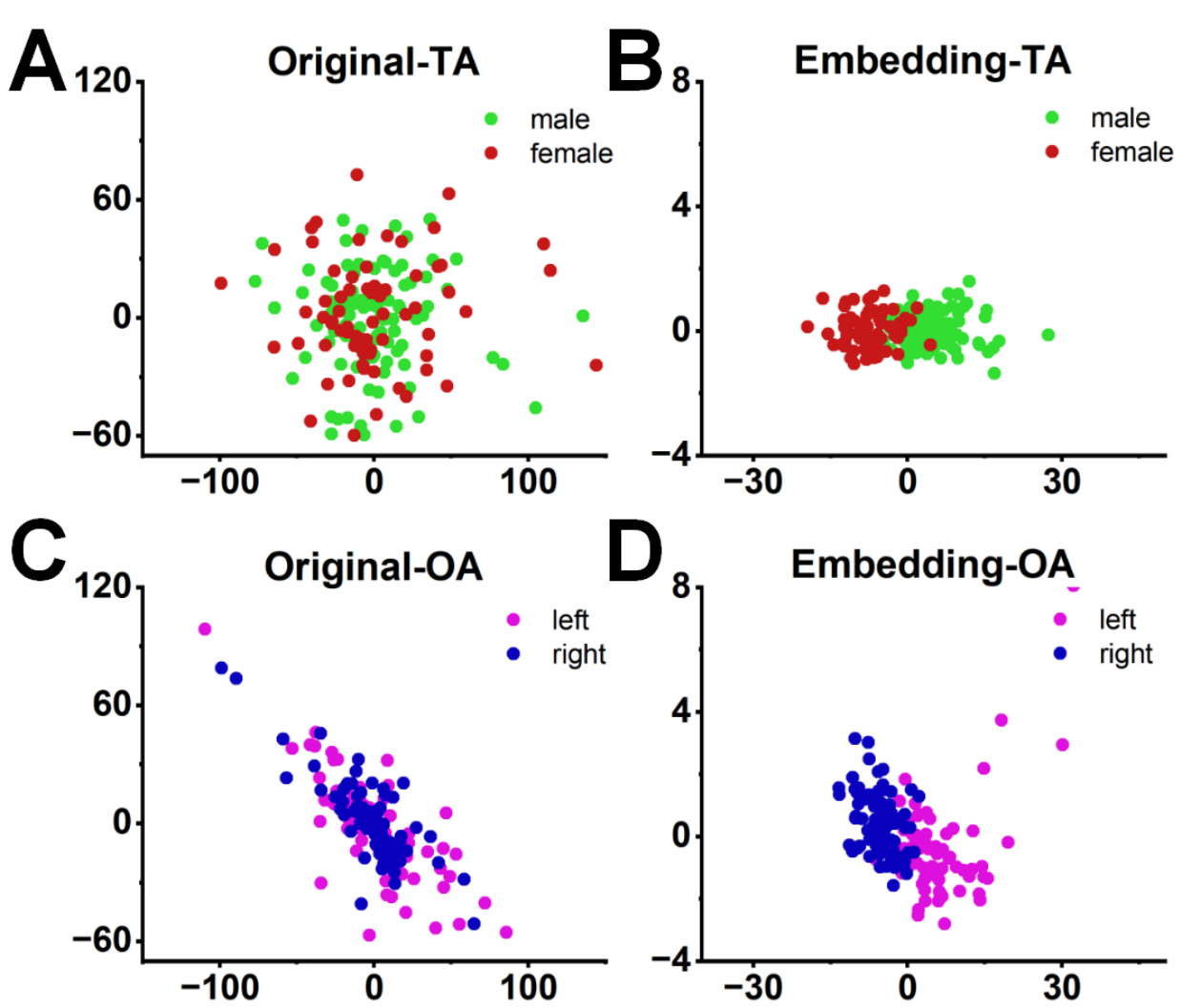

Fig. 4. Decision capability projection of AADNet in OA and TA tasks. (A) TA task, projection of the original EEG data distribution. (B) TA task, projection of features extracted by AADNet. (C) OA task, projection of the original EEG data distribution. (D) OA task, projection of features extracted by AADNet. The purpose of this figure is to demonstrate the effective decoding performance of the pretrained neural network employed in the study.

### D. Performance Evaluation

The decoding performance of three BCI paradigms with different channels was calculated with different strategies, including two RSs (averaging and voting) and one randomly selecting five sets of channels. For the two-class AAD task, ACC, area under the receiver operating characteristic curve (AUC), specificity (SPE), sensitivity (SEN), and F1 score, was employed for model performance evaluation. For the multiclass MI and AC tasks, only ACC was used as the evaluation metric. Besides, effective decoding ACC and computational efficiency (CE) was combined to select the optimal number of channels relevant to each task. CE is defined as the throughput of the neural network per second, measured in FPS (frames per second). The effective decoding ACC is defined as the decoding accuracy that must exceed the chance level for a specific task category.

## III Results

### A. Decoding performance with channel pruning

*a) AAD task*

The performance of AAD task with different number of the channels is displayed in Tables I and II. The decoding performance declines as the number of the channels decreasing under all the strategies, with a significant drop observed when channels are reduced from 10 to 5. Furthermore, compared to the random method, the two proposed RSs (averaging and voting) consistently demonstrated a significant advantage in mean decoding ACC across different channel counts, with this advantage becoming more pronounced as the number of channels decreased. Even with only 5 channels retained, the proposed RSs maintained decoding accuracy above 80% for the OA task and above 75% for the TA task, while the random method exhibited a further decrease of approximately 10%. Additionally, compared to the voting RS, the averaging RS demonstrated superior decoding accuracy, particularly when the channel count was 15.

Fig. 5 (A) and (B) clearly highlight the advantage of the averaging RS in decoding ACC and compare its stability with that of the random selection, to emphasize the effectiveness of the proposed approach. The results indicate that, in contrast to random channel selection, the averaging RS exhibits remarkable stability and reproducibility. Ultimately, due to its superior ability to maintain decoding performance, the averaging RS was chosen for channel pruning analysis in the AAD task.

*b) MI task*

The performance of the 4-class MI task with different channel counts is shown in Table III. Under all strategies, the decoding performance decreases as the number of channels reduces. Notably, when the number of selected channels was 5, the random method exhibited a significantly lower ability to select effective channels compared to the two proposed RSs. With the proposed RSs, the decoding ACC decreased by approximately 12%, whereas the random method saw a more substantial decline of about 19%. Additionally, the averaging RS demonstrated a more pronounced advantage in decoding performance compared to the voting RS.

Fig. 5 (C) more clearly illustrates the advantage of the averaging RS in maintaining decoding ACC and compares its stability with the random selection method. The results indicate that, compared to random channel selection, the averaging RS shows superior subject stability. Ultimately, due to its superior ability to maintain decoding performance, the averaging RS was applied in the MI task for channel pruning analysis.

*c) AC task*

In the AC task with more acquisition channels, the performance with different number of channels is shown in Table IV. Under all strategies, decoding performance decreased as the number of channels decreased. However, when the channel density $\eta$ was only 8%, all strategies still maintained a decoding accuracy of approximately 70%. Compared to the averaging RS, the voting RS demonstrated better retention of decoding accuracy, especially when the number of channels was 20, which differs from the results observed in the AAD and MI tasks. Nevertheless, when the channel density $\eta$ was less than 10%, the voting RS showed weaker ability to control the divergence in decoding accuracy compared to the averaging RS. Furthermore, although the random method showed slightly higher average decoding accuracy than the averaging RS, it was still inferior to the voting RS. Analysis of individual subject stability revealed that the performance repeatability of the random method was notably poor, with a maximum performance difference of over 20% across five random selections, as shown in Fig. 5 (D). Therefore, considering all factors, the voting RS was applied for channel pruning analysis in the AC task.

TABLE I
COMPARISON OF THE IMPACT OF THREE RANKING STRATEGIES AND CHANNEL SPARSITY PARAMETER $\eta$ ON OA DECODING PERFORMANCE. AT THE CHANNEL COUNT CORRESPONDING TO THE PERFORMANCE TURNING POINT, HIGHER PERFORMANCE IS HIGHLIGHTED IN BOLD.

| RS | Channel | | ACC (%) | AUC | F1 (%) | SPE (%) | SEN (%) |
|---|---|---|---|---|---|---|---|
| | $\eta$ | $C$ | | | | | |
| **Full** | 1.000 | 32 | 92.74(±3.43) | 0.976(±1.89%) | 92.89(±3.32) | 91.75(±5.61) | 93.68(±4.20) |
| **Avg** | 0.625 | 20 | 92.19(±4.62) | 0.972(±2.69%) | 92.20(±4.81) | 91.76(±5.78) | 92.44(±6.62) |
| | 0.469 | 15 | **92.02(±4.78)** | **0.970(±3.20%)** | **91.84(±5.34)** | **92.72(±4.25)** | 91.16(±7.78) |
| | 0.320 | 10 | 88.49(±5.30) | 0.950(±3.57%) | 88.40(±6.00) | 87.87(±7.92) | 88.89(±8.85) |
| | 0.156 | 5 | 80.50(±7.20) | 0.886(±6.14%) | 80.47(±7.56) | 79.19(±11.32) | 81.65(±10.57) |
| **Vote** | 0.625 | 20 | 92.42(±4.00) | 0.974(±2.18%) | 92.32(±4.37) | 92.49(±5.83) | 92.23(±5.47) |
| | 0.469 | 15 | 91.01(±4.25) | 0.967(±2.54%) | 90.97(±4.52) | 90.22(±5.75) | **91.69(±5.85)** |
| | 0.320 | 10 | 88.87(±5.20) | 0.952(±3.56%) | 88.63(±6.11) | 89.04(±7.65) | 88.39(±9.32) |
| | 0.156 | 5 | 80.99(±6.90) | 0.891(±6.07%) | 81.33(±7.29) | 77.20(±11.32) | 84.41(±9.99) |
| **Random** | 0.625 | 20 | 90.06(±4.04) | 0.960(±2.61%) | 90.04(±4.17) | 89.28(±5.79) | 90.75(5.76) |
| | 0.469 | 15 | 86.48(±5.33) | 0.934(±3.76%) | 86.34(±6.07) | 85.45(±7.06) | 87.30(±8.30) |
| | 0.320 | 10 | 81.37(±6.20) | 0.892(±5.22%) | 81.15(±7.23) | 79.29(±9.03) | 83.17(±9.91) |
| | 0.156 | 5 | 70.59(±6.69) | 0.781(±7.10%) | 69.19(±6.71) | 68.11(±11.28) | 72.44(11.28) |

TABLE II
COMPARISON OF THE IMPACT OF THREE RANKING STRATEGIES AND CHANNEL SPARSITY PARAMETER $\eta$ ON TA DECODING PERFORMANCE.

| RS | Channel | | ACC (%) | AUC | F1 (%) | SPE (%) | SEN (%) |
|---|---|---|---|---|---|---|---|
| | $\eta$ | $C$ | | | | | |
| **Full** | 1.000 | 32 | 90.04(±4.10) | 0.962(±2.38%) | 89.83(±4.25) | 88.49(±6.37) | 91.37(±6.85) |
| **Avg** | 0.625 | 20 | 89.45(±4.33) | 0.958(±2.50%) | 89.27(±4.62) | 87.65(±8.23) | 91.09(±7.74) |
| | 0.469 | 15 | **87.76 (±5.27)** | **0.945(±3.84%)** | **87.56(±5.52)** | **85.91(±9.48)** | 89.52(±8.79) |
| | 0.320 | 10 | 83.93(±6.20) | 0.916(±5.10%) | 83.59(±7.39) | 81.36(±11.77) | 86.08(±11.30) |
| | 0.156 | 5 | 76.19(±5.73) | 0.844(±6.19%) | 73.94(±10.10) | 76.11(±16.29) | 74.72(±17.43) |
| **Vote** | 0.625 | 20 | 88.79(±4.80) | 0.953(±3.17%) | 88.47(±5.38) | 87.49(±7.72) | 89.81(±8.46) |
| | 0.469 | 15 | 87.62(±5.34) | 0.943(±3.86%) | 87.53(±5.65) | 85.13(±8.66) | **89.83(±8.14)** |
| | 0.320 | 10 | 84.30(±6.15) | 0.918(±4.98%) | 83.93(±6.83) | 81.97(±11.70) | 86.14(±10.42) |
| | 0.156 | 5 | 75.07(±6.07) | 0.836(±5.85%) | 73.24(±10.20) | 73.31(±15.26) | 75.61(±15.94) |
| **Random** | 0.625 | 20 | 86.35(±4.47) | 0.936(±3.20%) | 86.03(±4.86) | 84.10(±7.59) | 88.28(±6.98) |
| | 0.469 | 15 | 82.36(±5.36) | 0.901(±4.75%) | 81.80(±6.39) | 79.97(±8.54) | 84.30(±9.36) |
| | 0.320 | 10 | 77.47(±5.31) | 0.857(±5.19%) | 76.01(±6.85) | 76.00(±9.56) | 78.06(±9.49) |
| | 0.156 | 5 | 65.68(±4.52) | 0.724(±5.75%) | 62.31(±10.53) | 63.08(±12.87) | 66.48(±12.87) |

TABLE III
THE IMPACT OF DIFFERENT RSS AND CHANNEL COUNTS ON MODEL DECODING ACCURACY IN THE MI TASK.

| Task | RS | Channel | | ACC (%) |
|---|---|---|---|---|
| | | $\eta$ | $C$ | |
| **MI** | **Full** | 1.000 | 22 | 80.75(±12.37) |
| | **Avg** | 0.68 | 15 | 79.17(±13.48) |
| | | 0.45 | 10 | **76.20(±14.00)** |
| | | 0.23 | 5 | 68.37(±13.15) |
| | **Vote** | 0.68 | 15 | 78.78(±12.98) |
| | | 0.45 | 10 | 75.62(±13.00) |
| | | 0.23 | 5 | 68.13(±14.98) |
| | **Random** | 0.68 | 15 | 77.89(±12.39) |
| | | 0.45 | 10 | 74.85(±12.84) |
| | | 0.23 | 5 | 61.54(±12.66) |

TABLE IV
THE IMPACT OF DIFFERENT RSS AND CHANNEL COUNTS ON MODEL DECODING ACCURACY IN THE AC TASK.

| Task | RS | Channel | | ACC (%) |
|---|---|---|---|---|
| | | $\eta$ | $C$ | |
| **AC** | **Full** | 1.000 | 62 | 88.11(±4.92) |
| | **Avg** | 0.65 | 40 | 85.42(±5.67) |
| | | 0.32 | 20 | 80.36(±5.87) |
| | | 0.16 | 10 | 75.74(±5.92) |
| | | 0.08 | 5 | 69.51(±5.33) |
| | **Vote** | 0.65 | 40 | 85.62(±5.12) |
| | | 0.32 | 20 | **81.39(±5.76)** |
| | | 0.16 | 10 | 75.79(±5.60) |
| | | 0.08 | 5 | 70.31(±6.21) |
| | **Random** | 0.65 | 40 | 85.55(±5.05) |
| | | 0.32 | 20 | 80.57(±5.93) |
| | | 0.16 | 10 | 75.90(±6.13) |
| | | 0.08 | 5 | 68.01(±4.94) |

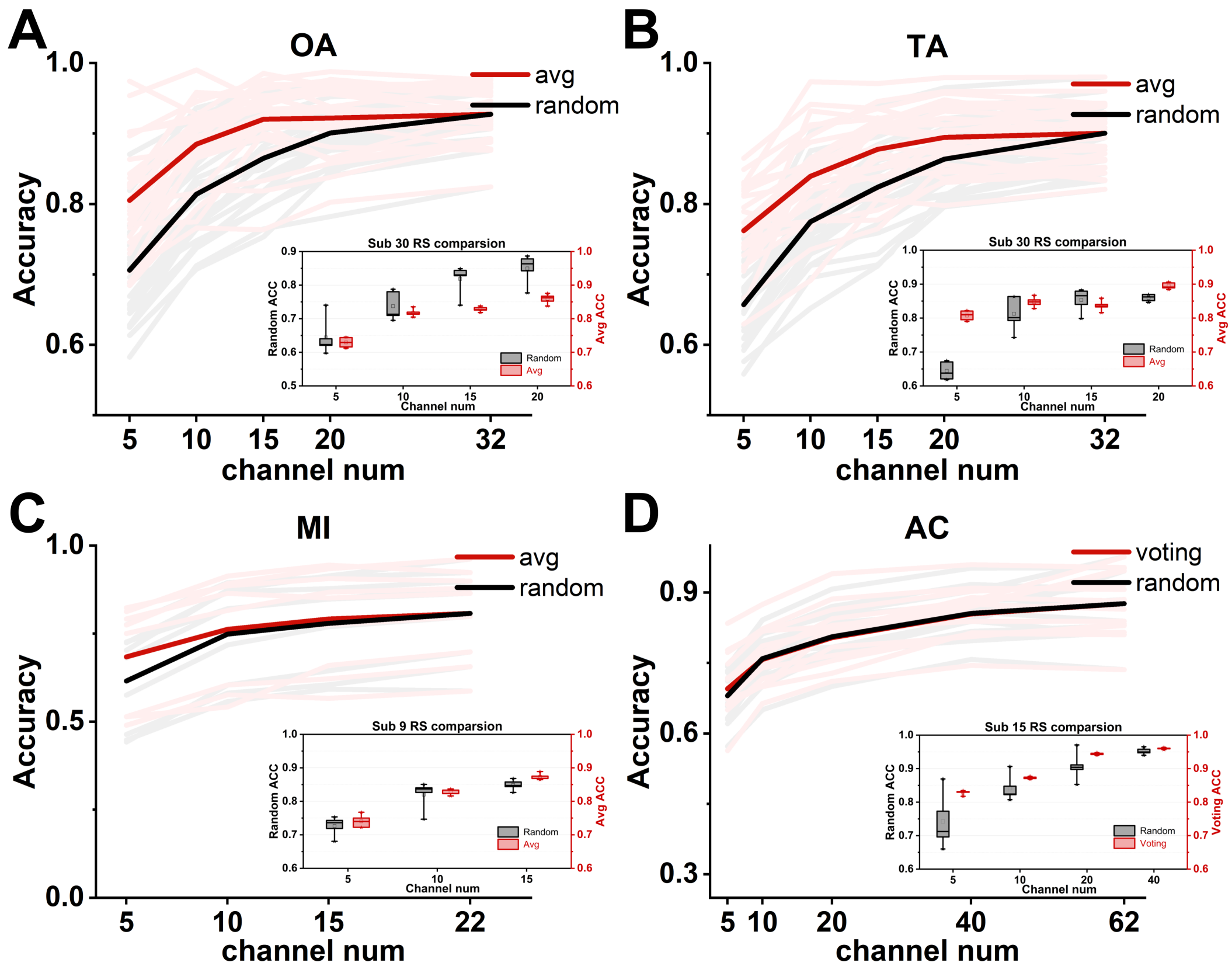

Fig. 5. The task decoding accuracy changes curves for all subjects at different channel counts (light lines), along with a comparison of the average performance of different strategies (dark lines). The small box plots represent the performance comparison between the selected RS strategies and the Random strategy for the last subject in each task. (A) OA task. (B) TA task. (C) MI task. (D) AC task. The aim is to demonstrate the positive impact of the proposed RS on decoding accuracy across different tasks, compared to random channel selection.

## B. Channel selection

Subject heterogeneity results in performance variability in BCI decoding task; however, the sensitivity to the optimal channel count remains relatively consistent across subjects. The decoding ACC variation curves for subjects in Fig. 5 indicate that, in the AAD task, the turning point for decoding accuracy change occurs at 15 channels; in the MI task, it occurs at 10 channels; and in the AC task, it occurs at 20 channels.

In addition, the balance curves between computational efficiency and decoding accuracy indicate that the optimal balance point for the AAD task is around 15 channels, for the MI task it is between 5 and 10 channels, and for the AC task it is around 10 channels, as shown in Fig. 6.

Taking into account factors such as decoding ACC, computational efficiency, electrode density, and positions (AAD task, Fig. 7; MI task, Fig. 8; AC task, Fig. 9), the final channel selection results are as follows: In the AAD task, considering the final OA and TA results, the selected channels $C_{AAD}$ are the top 15 electrodes ranked by classification contribution after RS ranking: Fp1, Fp2, F7, F8, AF4, AF3, F3, FC5, T8, T7, F4, CP6, C3, FC6, P8, and the electrode placement locations are shown in Fig. 7 (C); in the MI task, the selected channels $C_{MI}$ are the top 10 electrodes: C3, CP4, C4, CP3, CP2, C1, CPz, CP1, Cz, and POz, and the electrode placement locations are shown in Fig. 8 (B); and in the AC task, the selected channels $C_{AC}$ are the top 10 electrodes: T7, T8, TP7, FT7, FT8, C5, Oz, C1, P7, P5, the electrode placement locations are shown in Fig. 9 (B).

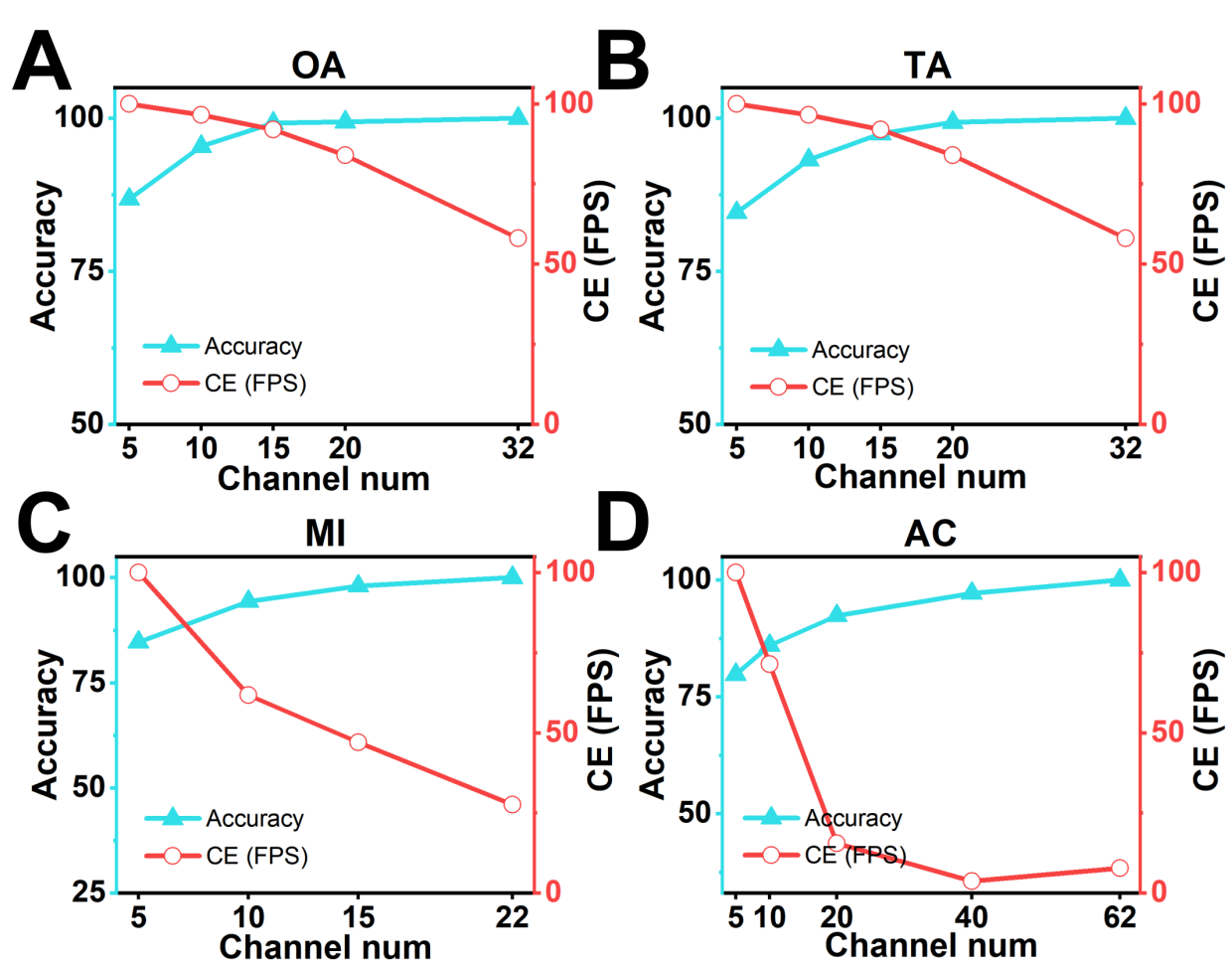

Fig. 6. Balance curves of computational efficiency and decoding accuracy for the AAD task (OA, TA), MI task, and AC task models. The left y-axis (in blue) represents the ACC at each channel count relative to the ACC with all channels, while the right y-axis (in red) represents the computational efficiency at each channel count relative to the maximum computational efficiency. (A) OA task, (B) TA task, (C) MI task, (D) AC task. The purpose of the figure is to illustrate the balance between decoding performance and computational efficiency, with the optimal channel count for achieving this balance varying across different tasks.

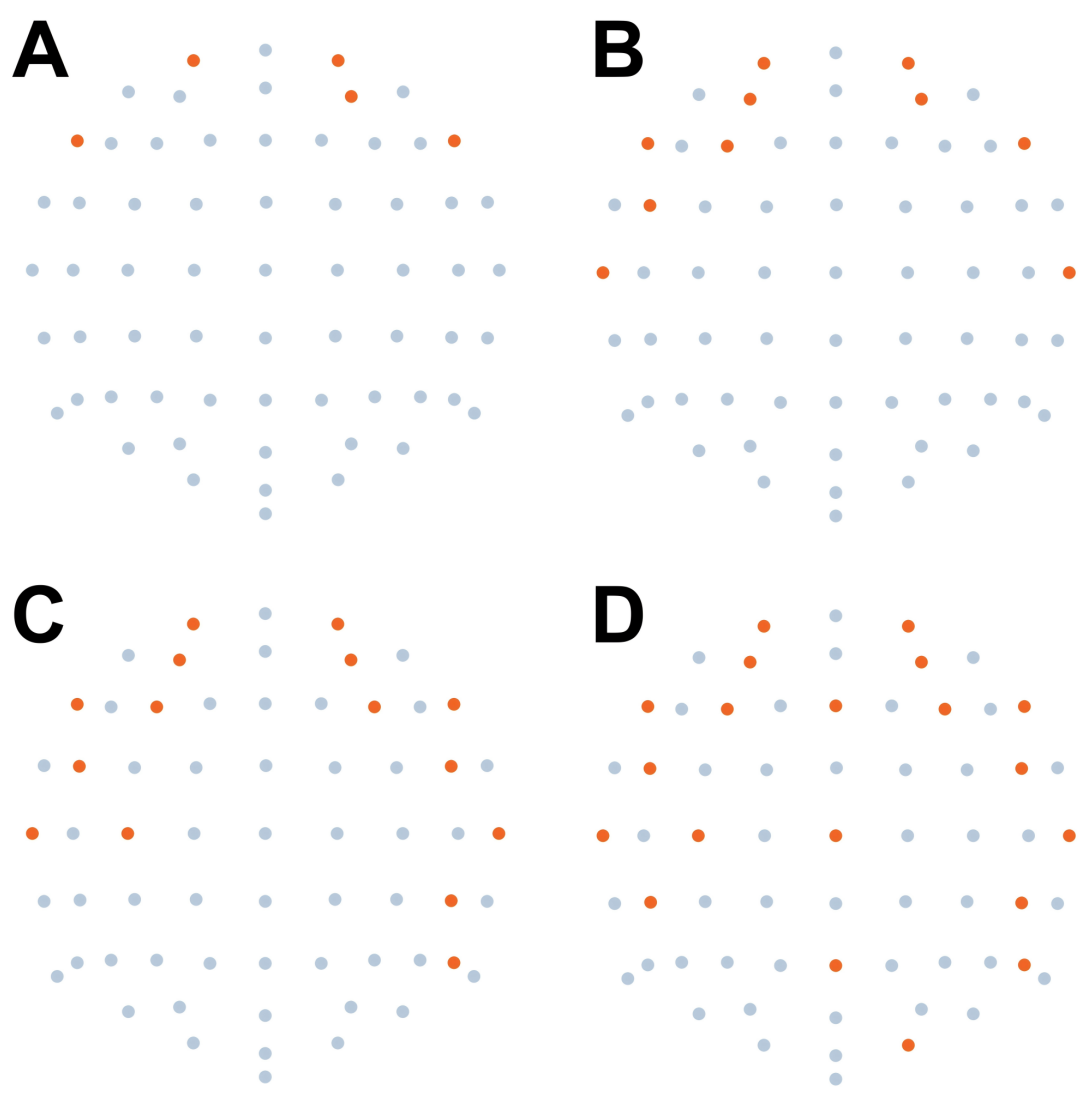

Fig. 7. AAD task channel selection results. Top (A) 5, (B) 10, (C) 15, (D) 20 channels in terms of scores. This figure aims to demonstrate the distribution of subsets of channels with varying densities selected by PlugSelect for the AAD task, highlighting its ability to identify electrophysiologically evoked channels related to auditory attention.

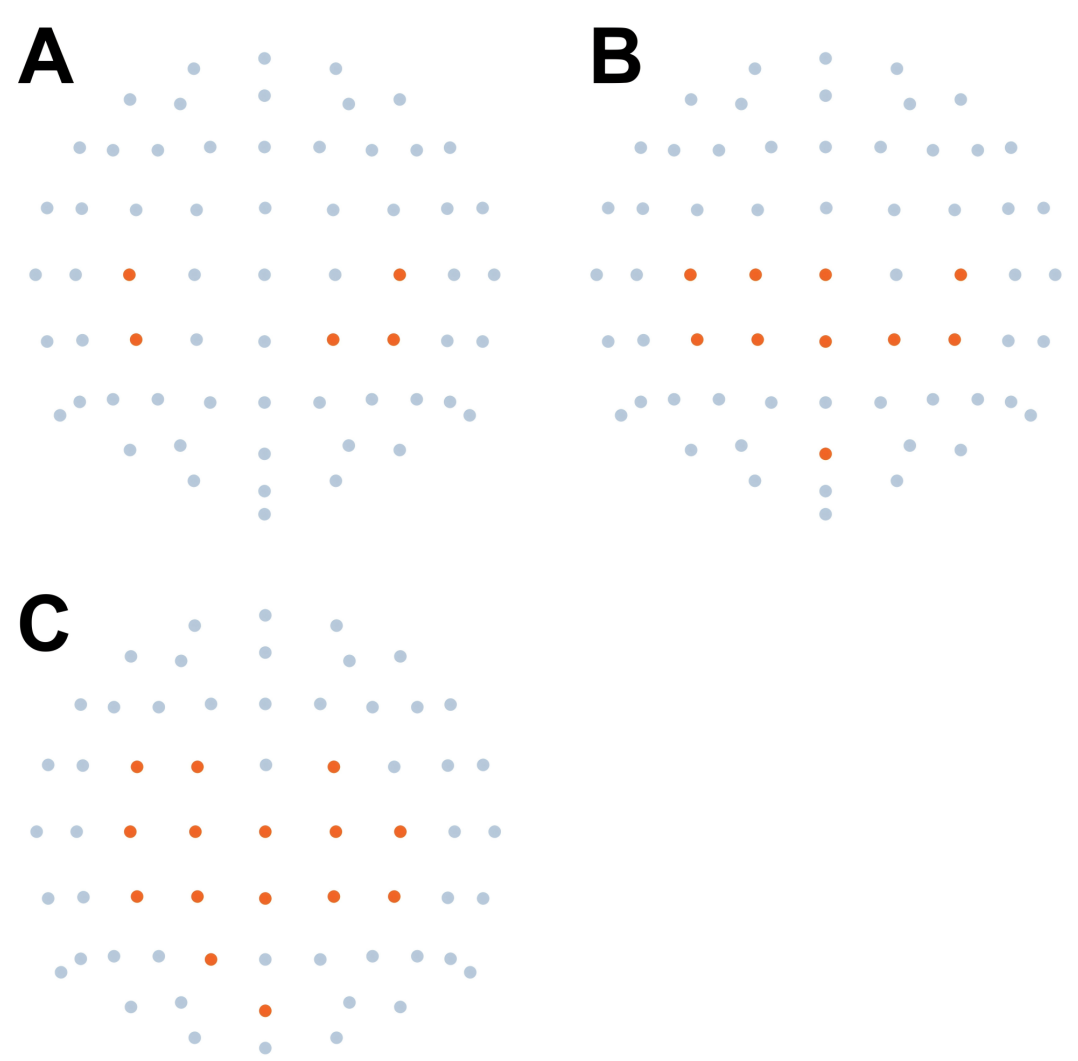

Fig. 8. MI task channel selection results. Top (A) 5, (B) 10, (C) 15 channels in terms of scores. This figure aims to illustrate the distribution of channel subsets with varying densities selected by PlugSelect for the four-class MI task, highlighting its ability to sensitively identify sensory-motor regions strongly associated with the task.

## IV Discussion

This study proposes PlugSelect, a channel selection framework for BCI tasks, aimed at practical applications such as portable neuro-steered hearing aids. The framework automatically selects channels through data-driven attentional weight assignment while ensuring high decoding efficiency. We validated and demonstrated the effectiveness of the proposed PlugSelect in channel selection, as well as its multi-platform portability and broad applicability, using the multi-attribute AAD dataset collected from 30 subjects, along with widely used MI (BCI Competition IV 2a) and AC datasets (SEED).

### *A. Plug-and-play PlugSelect, Multi-platform compatibility*

The main inference component of PlugSelect employs the IG algorithm to interpret the pre-trained model and automatically assign weights to each channel. Unlike end-to-end channel selection models [10], which add deployment and training costs, or channel selection modules [11] that may skew the learning direction of the classification model and reduce decoding performance, PlugSelect simplifies the process. And it avoids the issues associated with multiple iterations to find optimal channel combinations [8], which can increase noise interference and search difficulty. while channel sparsification methods like CSP reduce costs to some extent, they largely rely on prior selection or knowledge such as filters [15]. In contrast, PlugSelect streamlines system deployment by requiring only access to the pre-trained model and raw task data, and eliminates the need for classifier iteration or training, making it truly plug-and-play, as shown in Fig. 1. For this reason, as described in this paper, we can easily apply PlugSelect to BCI Competition IV-2a dataset, SEED, and other data platforms with different paradigms, categorization goals, and target quantities.

Additionally, due to individual differences among subjects, [25], [26] the optimal EEG channels vary from person to person (Fig. 5). PlugSelect can automatically select the most suitable sub-channel for each subject by using a subject-specific pre-training model, thus facilitating the personalized design of BCI devices for real-world applications.

### *B. Maintain decoding efficiency, Automate channel decoupling*

The optimal number of channels depends on the specific BCI paradigm and requirements. We also compared the impact of channel density on decoding efficiency. PlugSelect maintains a decoding efficiency similar to that of the full channel configuration when the channel density is greater than 65%, especially in the AAD task. When the number of channels is reduced from 32 to 15, TA decoding performance decreases by 2%, while OA performance decreases by less than 1%, and OA task performance remains more stable with a reduction in the number of channels. In the 4-class MI task, when the channel density is less than 50%, decoding accuracy decreases by less than 5%. In the AC task, when the channel density $\boldsymbol{\eta}$ is 0.32, emotional decoding accuracy remains above 80%. This indicates that PlugSelect can effectively reduce the number of channels unrelated to the task. Moreover, with this number of electrodes, mobile EEG recordings are feasible outside the laboratory and in everyday life settings [9].

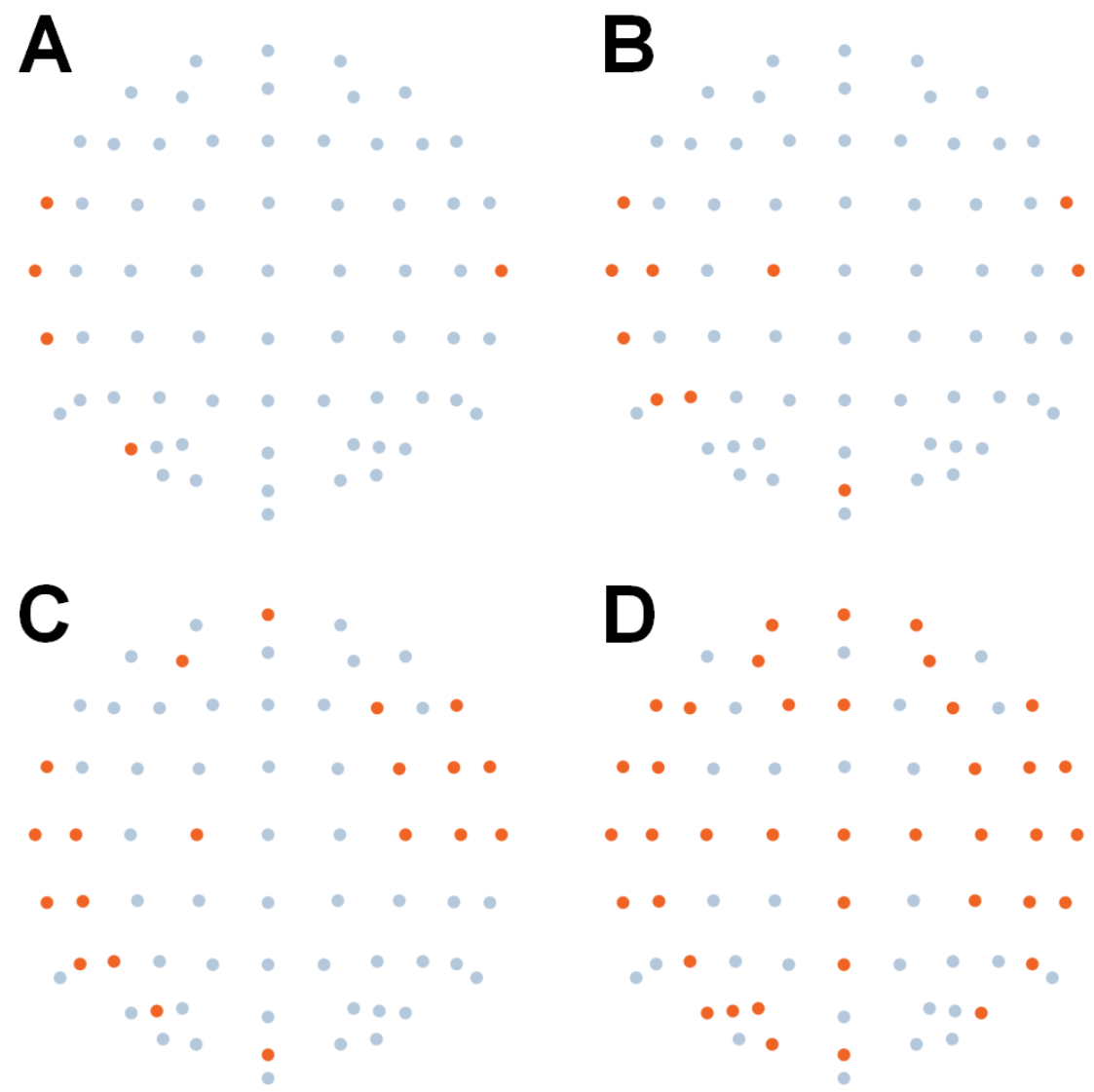

Fig. 9. AC task channel selection results. Top (A) 5, (B) 10, (C) 20, (D) 40 channels in terms of scores. The figure aims to illustrate the distribution of different channel density subsets selected by PlugSelect for the 3-class AC task, highlighting that the regions with high attribution identified among the 62 channels are primarily concentrated in the left and right temporal lobes.

Due to subject heterogeneity, we investigated the impact of different channel sequencing strategies on overall decoding performance to obtain a more stable RS and identify common activation patterns for specific BCI tasks. Firstly, compared to the random selection strategy, the two proposed RSs exhibit more stable performance and enhanced reproducibility. And Tables I, II, III, and IV indicate that the performance difference between the averaging strategy, which incorporates statistical sharing, and the voting strategy, which emphasizes individual differences, is around 1%. This suggests that PlugSelect's attribution calculation is insensitive to different ordering strategies and exhibits stronger stability. Additionally, we found that when the number of channels is larger, the voting RS outperforms the averaging RS. Notably, the decoding performance of subjects shows similar changes as the number of channels decreases, indicating that our model can effectively achieve automatic channel decoupling, identifying a set of invariant channels for specific BCI tasks and alleviating subject heterogeneity to some extent.

This evidence supports the notion that PlugSelect achieves channel decoupling and helps maintain decoding efficiency.

### *C. High relevance to downstream tasks*

PlugSelect explains the channel selection results by calculating the contribution levels. For the AAD task, the top 5 channels extracted by PlugSelect were Fp1, Fp2, F7, F8, and AF4, which are distributed within the prefrontal lobe. This is likely because the AAD task requires subjects to focus intently on attentional targets in a noisy environment, and the prefrontal lobe is linked to transient decision-making and attention allocation [27], [28].

The top three channels extracted by PlugSelect in the MI task were C3, CP4, and C4, which align with the findings of [7]. These channels are located in the primary sensorimotor cortex. Numerous EEG studies have confirmed that motor imagery activates primary sensorimotor areas [29], [30] and have observed significant ERD/ERS phenomena in the C3 and C4 regions [31].

In the AC task, the top five channels identified by PlugSelect were T7, T8, TP7, FT7 and FT8, consistent with the results from [32], [33], [34]. These channels are primarily situated in the temporal and frontal lobes. The temporal lobe is associated not only with auditory stimuli in this paradigm but also with audiovisual emotion comprehension [35]. And emotion cognition results from integrated processing across various brain regions [34], [36].

Moreover, since the multi-attribute AAD task paradigm used in this study is less commonly discussed, we can only analyze the plausibility of the channels extracted by PlugSelect from the perspective of functional brain regions. In contrast, the MI and ER paradigms and datasets are well-established, and the channels identified by PlugSelect are consistent with findings reported in existing literature, which serves as an additional validation of the reasonableness of the AAD channel results.

### *D. Research gap and future work*

While cross-subject automatic channel selection helps balance individual heterogeneity and group characteristics, this study primarily relied on statistical results from the best subchannels within single-subject domain to identify task-relevant invariant channels. In addition, the proposed PlugSelect relies on an effective pre-trained decoding model. Channels identified solely based on model inference may be limited in scope, and validating the effectiveness of the subset of channels extracted by PlugSelect through the functional properties of brain regions may lack sufficient accuracy. In future studies, we will further explore unsupervised algorithms to overcome labeling constraints and comprehensively evaluate the subset of channels associated with BCI task from multiple physiological perspectives.

## V Conclusion

In this study, we proposed a novel plug-and-play framework, PlugSelect, for efficient channel selection in BCI tasks. PlugSelect requires no additional training and can directly infer and interpret channels that are highly correlated with cortical electrical activity patterns through model result attribution. It is also efficiently portable across multiple platforms, such as AAD, MI and AC, addressing the limitations of existing algorithms. Furthermore, PlugSelect enables automatic channel decoupling while preserving decoding performance, providing a subset of channels closely related to specific BCI tasks, supporting the development and application of portable wearable devices.